# Compact Electrochromic Optical Recording of Bioelectric Potentials

Kenneth Nakasone,[1] Chris Zavik,[1] Erica Liu,[2,3] Burhan Ahmed,[1] Dana Griffith,[1] Lothar Maisenbacher,[1] Ashwin Singh,[1] Yuecheng Zhou,[2,3] Bianxiao Cui,[2,3] and Holger Müller[1]

[1] Department of Physics, 366 Physics South, University of California, Berkeley, CA 94720
[2] Department of Chemistry, 290 Jane Stanford Way, Stanford University, Stanford, CA 94305
[3] Wu Tsai Neurosciences Institute, 290 Jane Stanford Way, Stanford University, Stanford, CA 94305

**Abstract**

Electrochromic optical recording (ECORE) is a label-free method that utilizes electrochromism to optically detect electrical signals in biological cells with a high signal-to-noise ratio and is suitable for long-term recording. However, ECORE usually requires a large and intricate optical setup, making it relatively difficult to transport and to study specimens on a large scale. Here, we present a Compact ECORE (CECORE) apparatus that drastically reduces the spatial footprint and complexity of the ECORE setup whilst maintaining high sensitivity. An autobalancing differential photodetector automates common-mode noise rejection, removing the need for manually adjustable optics, and a compact laser module conserves space compared to a typical laser mount. The result is a simple, easy-to-use, and relatively low cost system that achieves a sensitivity of 16.7 µV (within a factor of 5 of the shot noise limit), and reliably detects action potentials from Human-induced pluripotent stem cell (HiPSC) derived cardiomyocytes. This setup can be further improved to within 1.5 dB of the shot noise limit by filtering out power-line interference.

**Introduction**

In-vitro electrophysiology has been used for drug testing [1,2], evaluation of stem-cell derived cardiomyocytes [3], and monitoring neuronal activities during long-term culture [4]. Different methods offer features such as minimal invasiveness, suitability for long-term recording, and scalability to multiple recording sites [5]. For example, the patch clamp is the gold standard for intracellular recording, but is invasive and limited to few recording sites and short recording durations. Multielectrode arrays (MEAs) enable multi-site recording [6] while optical recording offers flexibility in selecting target cells and high spatial resolution [7,8]. Genetically encoded fluorescent reporters can selectively monitor a desired subset of neurons and enable in-vivo voltage imaging [9], but have a slow response [10–12]. Voltage-dependent reporters such as potentiometric dyes [13] or fluorescent proteins [14,15] may have faster dynamics. However, at high frame rates (500–1000 frames per second), recordings are limited to seconds or minutes due to photobleaching [16], phototoxicity [17], or perturbations to the membrane capacitance [18,19]. Label-free optical methods [20] avoid these issues: Nitrogen-vacancy color centers in diamonds have been used to detect the magnetic field associated with electrical signals of excited worm and squid giant axons [21], but this requires averaging over numerous cells; electrically sensitive optical transitions in graphene have been used to detect spontaneous action potentials from whole chicken hearts, but without single-cell sensitivity[22].

Recently, we have demonstrated ElectroChromic Optical REcording (ECORE) which optically reads out bioelectrical signals by using the electrochromic effect [23] of π-conjugated polymer poly(3,4-ethylenedioxythiophene) poly(styrenesulfonate) (PEDOT:PSS) thin films. ECORE enables label-free, electrodeless long-term recordings with a high signal-to-noise ratio (SNR), avoiding photobleaching and phototoxicity. We have demonstrated ECORE-recording of spontaneous action potentials from individual cardiomyocytes and neurons, as well as cultured rat brain slices [24,25]. HiPSC-cardiomyocytes can be maintained on the PEDOT:PSS films for two months or longer [25].

ECORE, however, uses a balanced photodetector to cancel laser noise, which requires complexity to fine-tune the balance. As a result, existing ECORE setups are built on optical breadboards measuring half a meter and weighing tens of kilograms, making it relatively difficult to transport and study specimens at a large scale. Overcoming these limitations can make ECORE suitable for a wide range of application, such as use in incubators for culture screening, in-situ chemical sensing applications such as that of hydrogen peroxide produced by cancerous cells, or simultaneous use with other techniques to understand a variety of phenomena, such as cardiac proarrhythmia [26], deciphering the information encoded within neuronal networks [27], and developing brain-machine interfaces [28]. The compact and simple nature of compact ECORE directly translates into lower costs and makes use of the technique accessible more widely, such as for teaching in labs or routine screening of samples.

Here, we reduce the footprint and complexity of ECORE whilst maintaining performance. To achieve this, we radically simplify the optics by combining the focusing and routing functions into common optical elements and by using an autobalancing detector [29–31] to avoid the need to balance the laser powers. The simplicity and ease-of-use enables broad applications to detect and study bioelectric activities. We demonstrate a measurement of artificial electrical signals applied by a potentiostat as well as of bioelectric potentials in cardiomyocytes. Compact ECORE can fit into an incubator for long-term and high-throughput recording of cell action potentials.

**Methods**

Figure 1 shows a schematic diagram of compact ECORE. The number of optical elements has been reduced to seven (including the laser, the sample, and the detector), while maintaining the sensitivity that is necessary to view the biological potentials. As a light source, we use a 657-nm diode laser module (Thorlabs CPS650F) with a built-in adjustable lens that we use to focus the beam onto the sample surface to an 31 µm x 72 µm elliptical spot, which is roughly the size of a single cardiomyocyte [20]. The laser wavelength of this module (nominally 650 nm though ours was measured to emit at 657 nm) was chosen similar to the one used in Ref. [24] because of the availability of compact diode lasers; other wavelengths might lead to somewhat higher sensitivity [25]. A polarizing beamsplitter (PBS) cleans up the polarization. This strongly reduces technical noise by suppressing residual polarization fluctuations of the diode laser, which are converted into amplitude noise by the polarization-dependence of subsequent optics and the sample. The laser module is mounted on a rotation mount, which allows us to maximize the transmission through the PBS. A non-polarizing beamsplitter (BS) then separates the beam into two parts: a signal beam, which is directed towards the sample, and a reference beam for laser noise cancellation. An equilateral prism couples light to the sample and reflects it back out. Finally, an auto-balancing photodetector collects both the reference and signal beams.

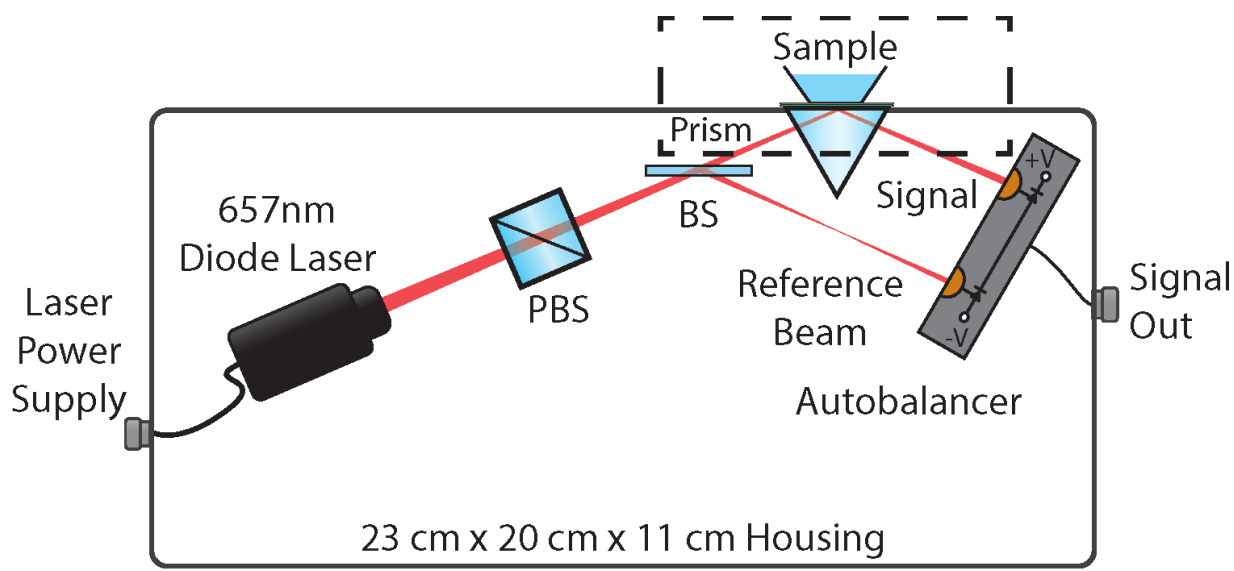

Fig. 1: Compact ECORE schematic. A commercial diode laser module (CPS650F, Thorlabs, Newton, NJ) with a built-in focusing lens emits a beam at 4.1 mW, which passes through a polarizing beam splitter (PBS) for polarization cleanup. A non-polarizing beam splitter (BS; Thorlabs BSX04) reflects ~80% of the laser power to form a reference beam that is sent directly to the reference photodiode of an autobalancing differential photodetector. The transmitted beam impinges on the sample via a 20-mm, equilateral prism out of BK-7 before being detected on the signal photodiode, traveling a total distance of roughly 12.5 cm. The dashed box highlights the sample-PEDOT-ITO-prism interface, which is shown in further detail in Fig. 2.

Since action potentials typically modulate the intensity of the laser beam by only about $10^{-3}$ or less, suppression of technical laser noise is crucial in ECORE to see this weak signal with a good signal to noise ratio[25]. Previously, this was achieved by detecting both the signal and the reference beam on two photodiodes and subtracting their photocurrents, which rejects common-mode noise, of which laser noise is one type [31]. The optical power of the two beams were manually balanced using a half-wave plate before a polarizing beamsplitter (wollaston prism). The balance needed to be adjusted to account for reflectivity changes due to variation in PEDOT film thickness across a single sample, as well as variations between samples, and required frequent manual intervention. This increases the number of necessary optical elements. In compact ECORE we make use of an autobalancing differential photodetector [29–31] in which the current from the reference photodiode is electronically attenuated by a variable ratio automatically, until it cancels the signal photocurrent. This automatically rejects technical laser noise without the need for manual intervention and with an extremely simple optical setup. The balance is adjusted slowly, with a time constant of 1.65 seconds, so as not to distort the fast action potentials.

## Results

*Cell-free measurement of applied electrical potentials*

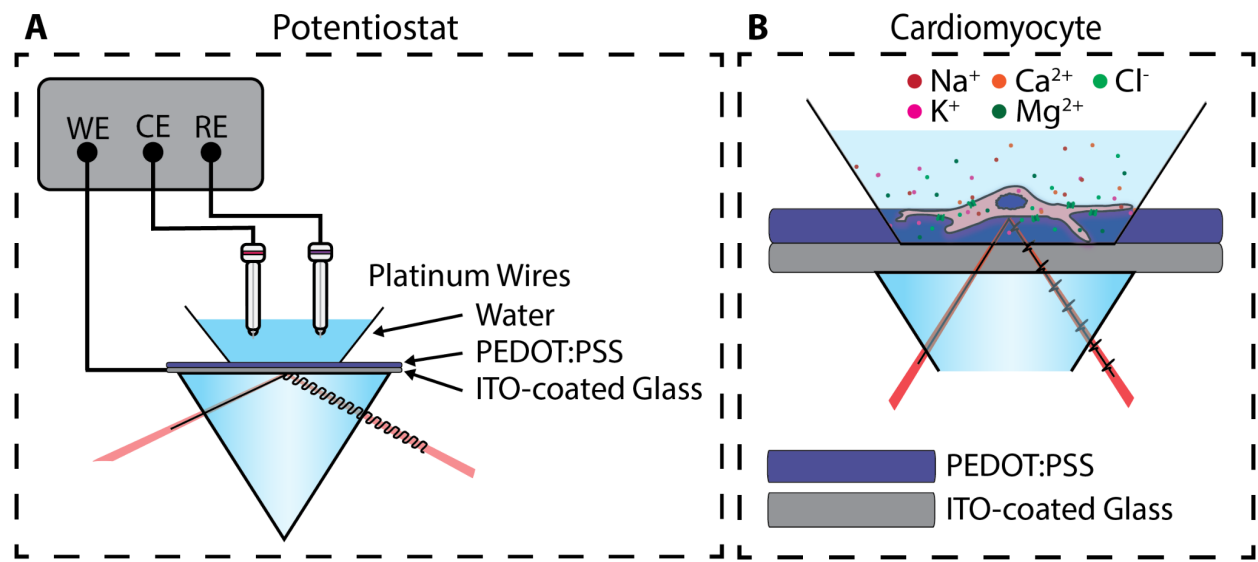

Fig. 2: (a) The setup for using CECORE to measure signals generated by a potentiostat, where WE is the working electrode, RE the reference electrode, and CE the counter electrode. (b) The setup for using CECORE to measure action potentials in cardiomyocytes.

Figure 2 shows how electrical signals are transduced into optical signals in ECORE. Light is reflected at the boundary between media with different complex refractive indices $n + i\kappa$, where $n \sim 1.4$ is the usual refractive index and $\kappa$ the extinction coefficient. At the open-circuit potential, $\kappa$ is typically $\sim 0.22$ and varies by about -0.57/V with applied voltage [24]. In Fig. 2a, the laser beam arrives from the left, traveling inside the BK-7 prism. The PEDOT film is mounted to a microscope slide and coupled to the prism via index-matching oil. The upper surface of the microscope slide is coated with indium-tin oxide (ITO) to form a working electrode, which in turn is coated with the ~ 50-nm thick PEDOT:PSS film. Electrical signals modulate the extinction coefficient of the PEDOT:PSS layer, which changes the reflectivity of the four-layer system (glass - ITO - PEDOT:PSS - water). The sensitivity is optimized near an incident angle of 67 degrees [24].

As a first demonstration, we used a potentiostat to apply a 1 mV, 10 Hz square wave and recorded the induced reflectivity changes via the output signal of the autobalancing photodetector. The recorded signal is shown on Fig. 3a. To characterize the SNR of this recording, we fit exponential functions to parts of the rising and falling sections of each charging-discharging cycle in the data, as shown in Fig. 3b. For each cycle, we determine the noise by calculating the standard deviation of the data from the exponential fits, while the peak-to-peak signal is determined by calculating the difference between the highest and lowest points of the exponential fits. We repeat this process for 250 cycles to give an average SNR of 60 for the 1 mV applied potential within a 1 kHz bandwidth. This is equivalent to a detection sensitivity of 16.7 µV, which is comparable to our previous measurements with ECORE and demonstrates that we are able to achieve a compact ECORE setup without sacrificing the sensitivity required to record cellular potentials. As in standard ECORE, these measurements in which signals are applied to the entire PEDOT surface exhibit a slower response time, limited by the bulk conductivity of the aqueous medium [24]. This is not a limitation for measurements with cells, in which potentials are applied only locally.

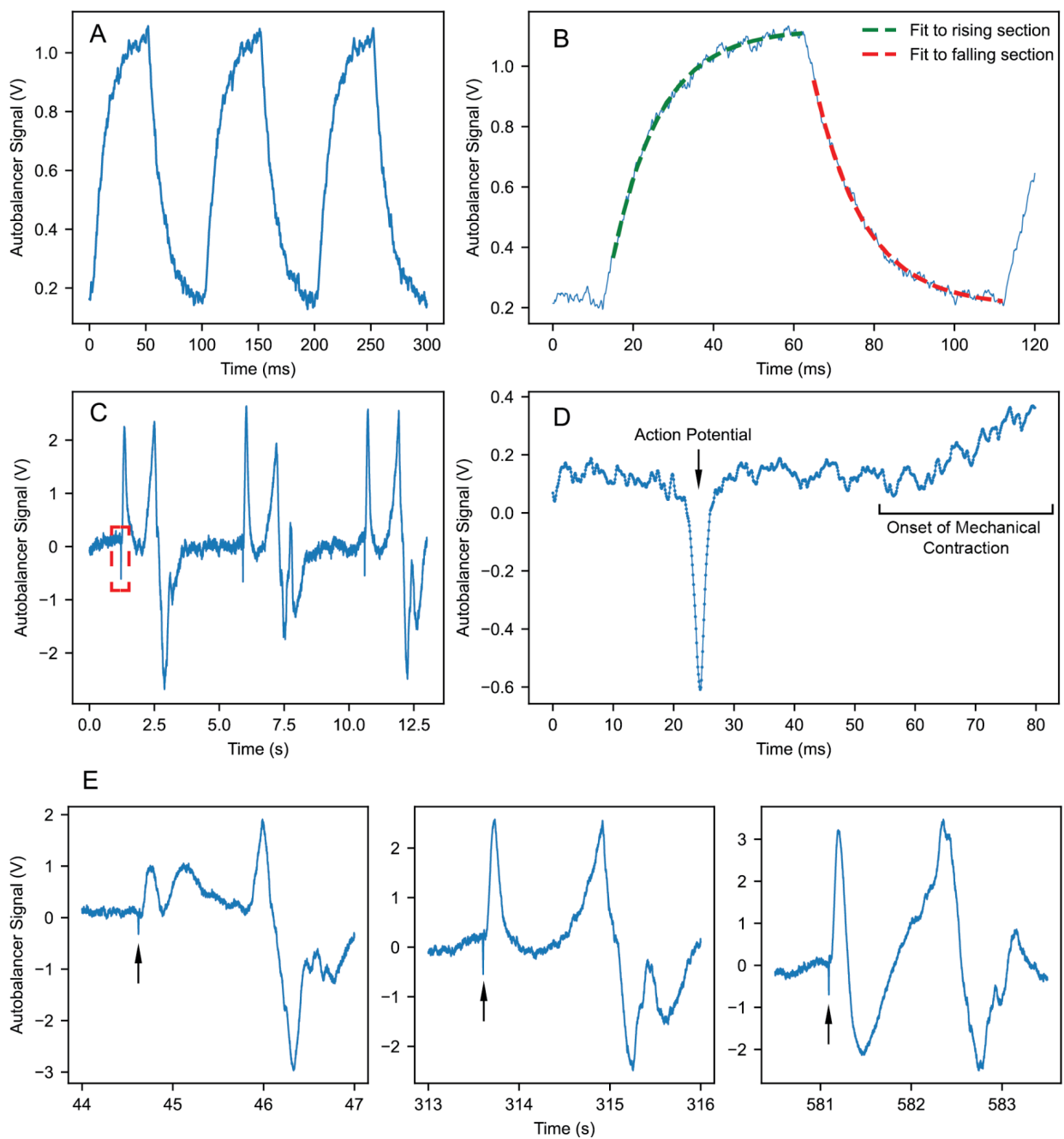

Fig. 3: Examples for recorded signals. Time values on the horizontal axes for Fig. 3a-3d have been translated to start at 0 s for ease of reading time intervals. (a): Cell-free measurement of a 1-mV peak-to-peak, 10-Hz square wave, applied via a potentiostat. 1 V in the plotted signal corresponds to a optical signal of 29 nW of the reflected beam. (b): Procedure for determining the SNR of CECORE. The solid blue line shows a typical portion of the recorded signal from the data depicted in Fig. 2a. The dashed lines show exponential fits to the data. The fits are used to determine both the noise and the peak-to-peak signal of CECORE. (c): Recording of human-induced pluripotent stem cell (iPSC)-derived cardiomyocytes, showing the electrical signal as a short spike, followed by a large signal resulting from the mechanical contraction of the cardiomyocytes. (d): Magnified plot of the boxed electrical signal from Fig. 3c. (e) A few samples of cardiomyocyte induced signals over the course of 9 minutes.

The SNR is ultimately limited by shot noise. Our autobalancing detector uses Thorlabs FDS100 photodiodes and a $R$ = 100 kΩ transfer impedance. The shot noise limit can be estimated from the detector output voltage $U$ as measured when the reference beam is blocked. The current spectral density is $\sqrt{2 \cdot 2eI} = \sqrt{4eU/R}$ for an output voltage of the detector $U$, where a factor of 2 has been inserted to account for the noise arising from the current from the reference photodiode, when it balances the current from the signal photodiode. This translates into a voltage spectral density of $\sqrt{4eUR}$ at the detector output, and thus into a root-mean-square (rms) noise of $\sqrt{4eURB}$ within a bandwidth $B$. For measurements like the ones shown in Fig. 3, we typically see $U \sim 0.13$ V and the equivalent noise bandwidth is $B$ = 1.08 kHz, which leads us to 3.0 µV$_{rms}$ of shot noise. The measurement shown in Fig. 3A shows about 15 µV$_{rms}$ of noise at the detector output, about a factor of 5 above shot noise. A Fourier transform of the noise floor shows that most of it is due to 60-Hz power line frequencies (Fig. 4). When these are removed, the noise floor is about 3.2 µV$_{rms}$ between 100 Hz and 1 kHz, which is within 1.5 dB of the shot noise limit.

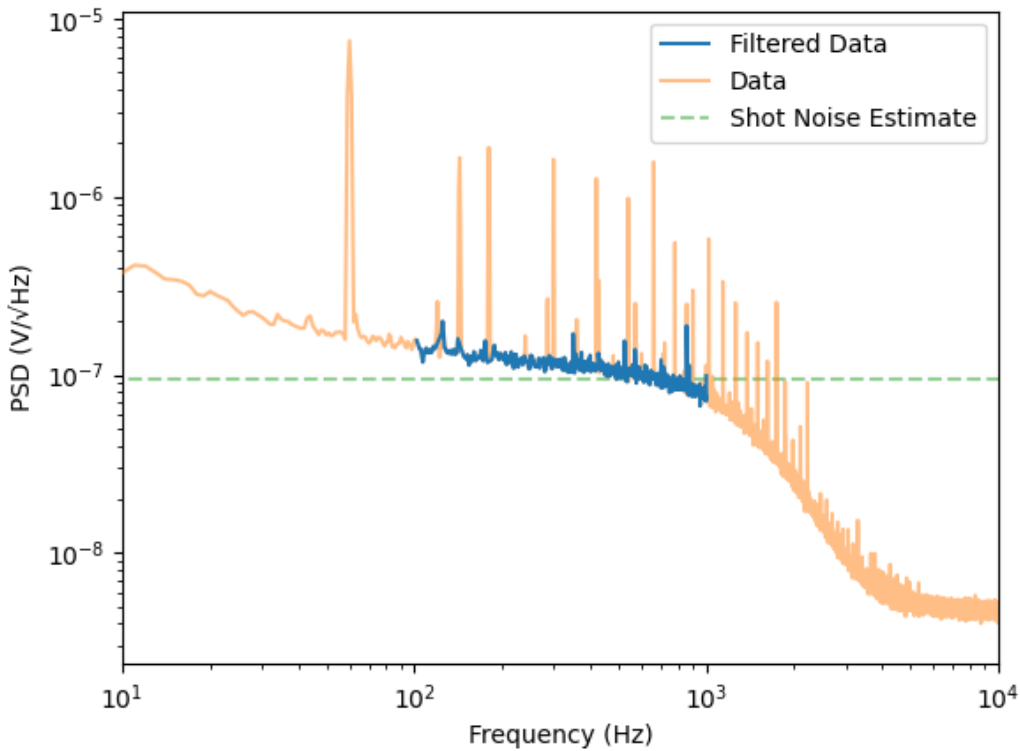

Fig. 4: Power spectral density of the noise floor at the detector output. The filtered data reflects the removal of periodic signals at 60Hz and 143Hz. The RMS of the filtered data in the plotted region is 3.20µV.

*Recordings with Cardiomyocytes*

In measurements involving cells, no electrodes are placed in the solution. Human-induced pluripotent stem cell (HiPSC)-derived cardiomyocytes were cultured on a 6-mm diameter PEDOT film for 9 days before the recording. The recorded signals are shown in Fig. 3c-3e. The electrical signal can be observed as a short spike. It is followed by a larger signal created by the mechanical contractions of cardiomyocytes. In our previous work [24,25], we have confirmed that the sharp spikes are, indeed, action potentials, by treating the cardiomyocytes with blebbistatin (12.5 µM) while the cells were being continuously recorded by ECORE. Blebbistatin is a potent myosin inhibitor that is known to inhibit cardiomyocyte contraction without eliminating its action potential [32]. Using compact ECORE, we detected the action potential shown in Fig. 3d with a signal-to-noise ratio of 38 and a half width of 0.97 ms. Even after 10 min continuous optical recording, the signal remains the same as shown in Figure 3e, confirming that compact ECORE is not limited by photobleaching or phototoxicity.

**Discussion**

We have described a compact (23 cm x 20 cm x 11 cm) ECORE and shown that it is able to observe electrical spikes by cardiomyocytes. The sensitivity, time resolution, and spatial resolution are similar to those of traditional ECORE, and are sufficient for single-cell recording. The noninvasive nature and the ability to select specific regions of interest on the sample are preserved. The small number of optical components makes it easier to adapt compact ECORE to environmental conditions (temperature and humidity).

**Acknowledgements**

The authors thank Garrett Louie, Cristian Panda, Pengwei Sun, and Matthew Tao for important discussions and assistance. This work was supported by the National Institutes of Health grant No. 1R01NS121934-01 and by the David and Lucile Packard Foundation (B.C. and H.M.)

# References

[1] O. Caspi, I. Itzhaki, I. Kehat, A. Gepstein, G. Arbel, I. Huber, J. Satin, and L. Gepstein, *In Vitro Electrophysiological Drug Testing Using Human Embryonic Stem Cell Derived Cardiomyocytes*, Stem Cells Dev. **18**, 161 (2009).
[2] M. G. Gunawan, S. S. Sangha, S. Shafaattalab, E. Lin, D. A. Heims-Waldron, V. J. Bezzerides, Z. Laksman, and G. F. Tibbits, *Drug Screening Platform Using Human Induced*

*Pluripotent Stem Cell-Derived Atrial Cardiomyocytes and Optical Mapping*, Stem Cells Transl. Med. **10**, 68 (2021).

[3] C. Prajapati, M. Ojala, H. Lappi, K. Aalto-Setälä, and M. Pekkanen-Mattila, *Electrophysiological Evaluation of Human Induced Pluripotent Stem Cell-Derived Cardiomyocytes Obtained by Different Methods*, Stem Cell Res. **51**, 102176 (2021).

[4] F. Seibertz et al., *Electrophysiological and Calcium-Handling Development during Long-Term Culture of Human-Induced Pluripotent Stem Cell-Derived Cardiomyocytes*, Basic Res. Cardiol. **118**, 14 (2023).

[5] L. Luan et al., *Recent Advances in Electrical Neural Interface Engineering: Minimal Invasiveness, Longevity, and Scalability*, Neuron **108**, 302 (2020).

[6] Z. Zhao, H. Zhu, X. Li, L. Sun, F. He, J. E. Chung, D. F. Liu, L. Frank, L. Luan, and C. Xie, *Ultraflexible Electrode Arrays for Months-Long High-Density Electrophysiological Mapping of Thousands of Neurons in Rodents*, Nat Biomed Eng **7**, 520 (2023).

[7] M. Scanziani and M. Häusser, *Electrophysiology in the Age of Light*, Nature **461**, 930 (2009).

[8] B. A. Wilt, L. D. Burns, E. T. Wei Ho, K. K. Ghosh, E. A. Mukamel, and M. J. Schnitzer, *Advances in Light Microscopy for Neuroscience*, Annu. Rev. Neurosci. **32**, 435 (2009).

[9] H. H. Yang, F. St-Pierre, X. Sun, X. Ding, M. Z. Lin, and T. R. Clandinin, *Subcellular Imaging of Voltage and Calcium Signals Reveals Neural Processing In Vivo*, Cell **166**, 245 (2016).

[10] L. Tian et al., *Imaging Neural Activity in Worms, Flies and Mice with Improved GCaMP Calcium Indicators*, Nat. Methods **6**, 875 (2009).

[11] C. Grienberger and A. Konnerth, *Imaging Calcium in Neurons*, Neuron **73**, 862 (2012).

[12] T.-W. Chen et al., *Ultrasensitive Fluorescent Proteins for Imaging Neuronal Activity*, Nature **499**, 295 (2013).

[13] E. W. Miller, *Small Molecule Fluorescent Voltage Indicators for Studying Membrane Potential*, Curr. Opin. Chem. Biol. **33**, 74 (2016).

[14] H. H. Yang and F. St-Pierre, *Genetically Encoded Voltage Indicators: Opportunities and Challenges*, J. Neurosci. **36**, 9977 (2016).

[15] Y. Xu, P. Zou, and A. E. Cohen, *Voltage Imaging with Genetically Encoded Indicators*, Curr. Opin. Chem. Biol. **39**, 1 (2017).

[16] T. H. Grandy, S. A. Greenfield, and I. M. Devonshire, *An Evaluation of in Vivo Voltage-Sensitive Dyes: Pharmacological Side Effects and Signal-to-Noise Ratios after Effective Removal of Brain-Pulsation Artifacts*, J. Neurophysiol. **108**, 2931 (2012).

[17] H. Hirase, V. Nikolenko, and R. Yuste, *Multiphoton Stimulation of Neurons and Spines*, Cold Spring Harb. Protoc. **2012**, 472 (2012).

[18] S. Mennerick, M. Chisari, H.-J. Shu, A. Taylor, M. Vasek, L. N. Eisenman, and C. F. Zorumski, *Diverse Voltage-Sensitive Dyes Modulate GABAAReceptor Function*, J. Neurosci. **30**, 2871 (2010).

[19] D. Maclaurin, V. Venkatachalam, H. Lee, and A. E. Cohen, *Mechanism of Voltage-Sensitive Fluorescence in a Microbial Rhodopsin*, Proc. Natl. Acad. Sci. U. S. A. **110**, 5939 (2013).

[20] Y. Zhou, E. Liu, H. Müller, and B. Cui, *Optical Electrophysiology: Toward the Goal of Label-Free Voltage Imaging*, J. Am. Chem. Soc. **143**, 10482 (2021).

[21] J. F. Barry, M. J. Turner, J. M. Schloss, D. R. Glenn, Y. Song, M. D. Lukin, H. Park, and R. L. Walsworth, *Optical Magnetic Detection of Single-Neuron Action Potentials Using Quantum Defects in Diamond*, Proceedings of the National Academy of Sciences **113**, 14133 (2016).

[22] H. B. Balch, A. F. McGuire, J. Horng, H.-Z. Tsai, K. K. Qi, Y.-S. Duh, P. R. Forrester, M. F. Crommie, B. Cui, and F. Wang, *Graphene Electric Field Sensor Enables Single Shot Label-Free Imaging of Bioelectric Potentials*, Nano Lett. **21**, 4944 (2021).

[23] P. M. Beaujuge and J. R. Reynolds, *Color Control in Pi-Conjugated Organic Polymers for Use in Electrochromic Devices*, Chem. Rev. **110**, 268 (2010).

[24] F. S. Alfonso et al., *Label-Free Optical Detection of Bioelectric Potentials Using Electrochromic Thin Films*, Proc. Natl. Acad. Sci. U. S. A. **117**, 17260 (2020).

[25] Y. Zhou, E. Liu, Y. Yang, F. S. Alfonso, B. Ahmed, K. Nakasone, C. Forró, H. Müller, and B. Cui, *Dual-Color Optical Recording of Bioelectric Potentials by Polymer Electrochromism*, J. Am. Chem. Soc. **144**, 23505 (2022).

[26] T. Colatsky, B. Fermini, G. Gintant, J. B. Pierson, P. Sager, Y. Sekino, D. G. Strauss, and N. Stockbridge, *The Comprehensive in Vitro Proarrhythmia Assay (CiPA) Initiative - Update on Progress*, J. Pharmacol. Toxicol. Methods **81**, 15 (2016).

[27] Y. Adam et al., *Voltage Imaging and Optogenetics Reveal Behaviour-Dependent Changes in Hippocampal Dynamics*, Nature **569**, 413 (2019).

[28] E. Musk and Neuralink, *An Integrated Brain-Machine Interface Platform With Thousands of Channels*, J. Med. Internet Res. **21**, e16194 (2019).

[29] P. C. Hobbs, *Ultrasensitive Laser Measurements without Tears*, Appl. Opt. **36**, 903 (1997).

[30] K. L. Haller and P. C. D. Hobbs, *Double-Beam Laser Absorption Spectroscopy: Shot Noise-Limited Performance at Baseband with a Novel Electronic Noise Canceler*, in *Optical Methods for Ultrasensitive Detection and Analysis: Techniques and Applications*, Vol. 1435 (SPIE, 1991), pp. 298–309.

[31] P. C. D. Hobbs, *Building Electro-Optical Systems: Making It All Work* (John Wiley & Sons, 2011).

[32] V. V. Fedorov, I. T. Lozinsky, E. A. Sosunov, E. P. Anyukhovsky, M. R. Rosen, C. W. Balke, and I. R. Efimov, *Application of Blebbistatin as an Excitation–contraction Uncoupler for Electrophysiologic Study of Rat and Rabbit Hearts*, Heart Rhythm **4**, 619 (2007).